\documentclass{aa}
\usepackage[varg]{txfonts}
\usepackage{graphicx}

\usepackage{natbib}
\bibpunct{(}{)}{;}{a}{}{,}             

\newcommand\kms{$\mbox{km s}^{-1}$}

\newcommand\z{$\phantom{0}$}

\begin{document}

\title{Multiplicity of the red supergiant population in the young massive cluster NGC\,330\thanks{Based on observations collected at the European Organisation for Astronomical Research in the Southern Hemisphere under ESO programmes 60.A-9183, 083.C-0413, 083.D-0549, 084.D-0591, 085.C-0614, 085.D-0395, 086.D-0078. Table~\ref{tb:rvall} is only available in full in electronic form at the CDS via anonymous ftp to cdsarc.u-strasbg.fr (130.79.128.5)
or via http://cdsweb.u-strasbg.fr/cgi-bin/qcat?J/A+A/}}
\titlerunning{Multiplicity of red supergiants in NGC\,330}

\author{L. R. Patrick\inst{1, 2}
  \and D. J. Lennon\inst{1, 2}
  \and C. J. Evans\inst{3}
  \and H. Sana\inst{4}
  \and J. Bodensteiner\inst{4}
  \and N. Britavskiy\inst{1, 2}
  \and R. Dorda\inst{1, 2}
  \and\\ A. Herrero\inst{1, 2}
  \and I. Negueruela\inst{5}
  \and A. de Koter\inst{4, 6}
     }

\offprints{lpatrick@iac.es}

\institute{Instituto de Astrof\'isica de Canarias, E-38205 La Laguna, Tenerife, Spain \email{lpatrick@iac.es}
  \and Universidad de La Laguna, Dpto. Astrof\'isica, E-38206 La Laguna, Tenerife, Spain
  \and UK Astronomy Technology Centre, Royal Observatory, Blackford Hill, Edinburgh, EH9 3HJ, UK
  \and Institute of astrophysics, KU Leuven, Celestijnlaan 200D, 3001, Leuven, Belgium
  \and Departamento de F\'isica Aplicada, Facultad de Ciencias, Universidad de Alicante, Carretera San Vicente s/n, E03690, San Vicentedel Raspeig, Spain
  \and Anton Pannenkoek Institute for Astronomy, Universiteit van Amsterdam, Science Park 904, NL-1098 XH Amsterdam, The Netherlands
}

\date{Received September 2019}

\abstract{The multiplicity properties of massive stars are one of the
  important outstanding issues in stellar evolution. Quantifying the
  binary statistics of all evolutionary phases is essential to paint a
  complete picture of how and when massive stars interact with their
  companions, and to determine the consequences of these interactions.}
  {We investigate the multiplicity of an almost complete census of red
  supergiant stars (RSGs) in NGC\,330, a young massive cluster in the Small
  Magellanic Cloud.}  {Using a combination of multi-epoch HARPS and MUSE
  spectroscopy, we estimate radial velocities and assess the
  kinematic and multiplicity properties of 15 RSGs in
  NGC\,330.}  {Radial velocities are
  estimated to better than $\pm$100\,m\,s$^{-1}$ for the HARPS data.
  The line-of-sight velocity dispersion for the cluster is estimated
  as $\sigma_{\rm 1D}$\,$=$\,3.20\,$^{+0.69}_{-0.52}$\,\kms.
  When virial equilibrium is assumed, the dynamical mass of the cluster is
  log\,(M$_{\rm dyn}$/M$_{\odot}$)\,$=$\,5.20\,$\pm$\,0.17, in good agreement with
  previous upper limits.  We detect significant radial velocity
  variability in our multi-epoch observations and distinguish between
  variations  caused by atmospheric activity and those caused by binarity. The
  binary fraction of NGC\,330 RSGs is estimated by comparisons with
  simulated observations of systems with a range of input binary
  fractions. In this way, we account for observational biases and
  estimate the intrinsic binary fraction for RSGs in NGC\,330 as
  $f_{\rm RSG}$~=~0.3\,$\pm$\,0.1 for orbital periods in the range
  2.3\,$<$\,log\,P\,[days]\,$<$\,4.3, with $q >$~0.1.
  Using the distribution of the luminosities of the RSG population, we estimate the age of NGC\,330 to be 45\,$\pm$\,5\,Myr and estimate a red straggler fraction of 50\%.}
  {We estimate the binary fraction of RSGs in NGC\,330 and conclude that it appears to be lower than that of main-sequence massive stars, which is expected because interactions between an RSG and a companion are assumed to effectively strip the RSG envelope.}


\maketitle

\section{Introduction}     \label{sec:introduction}

Most massive stars ($>$8\,M$_\odot$) reside in binary or higher order multiple
systems~\citep[e.g.][]{2013A&A...550A.107S,2014ApJS..215...15S,2015A&A...580A..93D,2017ApJS..230...15M}, and
$\sim$70\% interact with a companion during their
lifetimes~\citep{2012Sci...337..444S,2014ApJS..213...34K}.  These
interactions have profound effects on the evolution of all stars involved~\citep{2013ApJ...764..166D} and the nature of their subsequent supernova explosions~\citep{1992ApJ...391..246P,2017PASA...34....1D}.

Red supergiant stars (RSGs) are the evolved products of massive
main-sequence (OB-type) stars with initial masses in the range
8\,$<$\,M\,$<$\,40\,M$_{\odot}$~\citep[e.g.][]{2012A&A...537A.146E}.
The most numerous core-collapse supernovae by type are those classified as type
II-P, which are thought to arise from stars with initial masses of
8\,$<$\,M\,$<$\,23\,M$_{\odot}$ that explode while in the RSG phase
\citep[][although see \citet{2018MNRAS.474.2116D} for a
re-evaluation of the upper mass limit.]{Smartt09}.
The most massive RSGs are
thought to evolve back to hotter temperatures and likely explode as
blue supergiant stars~\citep[e.g. SN1987A,][]{1987ApJ...323L..35S}.
The RSG phase of evolution therefore is an important factor in the
yields of core-collapse supernovae in
general~\citep{Smartt09,2015PASA...32...16S}, and both the binary
fraction and multiplicity properties of this evolutionary stage have a
strong effect.

As binary systems evolve, the interactions between companions alter not
only the evolution of the individual stars, but also affect
the measured properties of the population.  Stars with close
companions merge and rejuvenate, spending more time on the main
sequence as blue straggler stars~\citep[e.g.][]{1964MNRAS.128..147M,2014ApJ...780..117S}.  When blue
straggler stars evolve to the RSG phase, this likely produces a
red straggler effect~\citep{b19}.

Massive main-sequence stars with companions of intermediate separation
and periods~\citep[between around 10 up to 1500\,days;][]{1992ApJ...391..246P,2012Sci...337..444S} interact as the primary evolves off the main sequence and begins to
expand dramatically.  Binary stellar evolutionary models predict that
interacting companions shorten the RSG lifetime by a factor of three
at solar metallicities and even more at lower metallicities,
resulting in hot massive
stars~\citep{2008MNRAS.384.1109E,2018A&A...615A..78G}, that is, fewer RSGs.

The remaining massive main-sequence stars in binary systems in which the separations of the
companions are sufficiently large to prevent significant
interaction until at least one companion reaches the RSG
phase can in principle be observed through various different
methods~\citep[e.g.][]{2018AJ....156..225N,2019ApJ...875..124N,p19}.

Placing observed binary systems in the context of stellar evolution
requires careful consideration of the inherent biases of
observations and the parameter space over which binary fraction
estimates are valid. 
By measuring radial velocity (RV) variations (or the lack thereof), we are able to
estimate the binary fraction and characterise the observed or excluded
systems.  Given the arguments outlined above, the
overall binary fraction of RSGs is expected to be significantly smaller than that
of main-sequence massive stars.

To date, relatively few Galactic RSGs are known to reside in binary
systems~\citep[e.g. VV Cep,][]{1977JRASC..71..152W}. The most easily
detectable configuration for an RSG in a binary system is with a B-type
companion, and this is indeed mainly what is observed in the Galaxy.
To find such systems, \citet{2018AJ....156..225N} defined photometric
criteria to identify RSGs that are contaminated by blue light from a
potential companion. These were recently expanded upon by spectroscopic follow-up of
potential binaries in M31, M33, and the Small Magellanic Cloud (SMC) by \citet{2019ApJ...875..124N}, finding that many of the sources with such a blue excess also show spectroscopic evidence of this crowding, which these authors interpret as evidence for binarity.
A complementary approach for detecting binarity is
via multi-epoch spectroscopy of RSGs.
\citet[][hereafter P19]{p19} provided the first estimate of the RSG binary fraction
($f_{\rm RSG}$) in the 30~Doradus region of the Large Magellanic Cloud
(LMC). For orbital periods in the
range 3.3\,$<$\,log\,P\,[days]\,$<$\,4.3, they found $f_{\rm
  RSG}$\,$\sim$\,0.3, in good agreement with
expectations~\citep{2017ApJS..230...15M}.

Young massive star clusters are the perfect environment in which to
hunt for RSG binaries~\citep{2009MNRAS.400.1479S,2012ApJ...751....4K}.
The Local Group of galaxies contains many well-catalogued young
massive clusters that contain significant populations of RSGs,  for
instance, h- and $\chi$-Persei \citep{2014ApJ...788...58G}, RSGC01
\citep{2008ApJ...676.1016D}, RSGC02 \citep{2007ApJ...671..781D}, and
RSGC03 \citep{2009A&A...498..109C} in the Galaxy, and
NGC\,2100~\citep{2016MNRAS.458.3968P} and Hodge\,301 (P19) in the LMC.
At the lower metallicity
of the SMC, the young massive cluster NGC\,330 has a well-studied
population of RSGs~\citep{1959AJ.....64..254A,1974A&AS...15..261R,1980MNRAS.191..285F} and
early-type massive stars
\citep{f72,g92,1996A&A...314..243L,2003A&A...398..455L,2006A&A...456..623E}.
By targeting such clusters with long-baseline multi-epoch
spectroscopic campaigns, we can begin to unveil the binary population
of RSGs.

Given the large radii of RSGs, orbital periods shorter than several
hundred days for a binary companion cannot be supported (P19).
Therefore, hunting for characteristic periodic RV
variations is a long-term endeavour as most variations arising from
binarity are expected to be of the order of 1-5\,\kms, on timescales
of several years or more. This amplitude and timescale is also
comparable to the variations seen from convective motions in the
atmospheres of RSGs, so that care must be given to distinguishing them from
genuine binarity.

Red supergiants display a dense forest of stellar absorption features at visible
wavelengths that can be useful for high-resolution chemical abundance
studies~\citep[e.g.][]{Cunha07}, but the blending of these features can be a
complicating factor at lower resolution~\citep[see ][for a recent
analysis of atomic line diagnostics]{2019arXiv190201862D}.  P19
developed a method to exploit this dense forest of absorption features
to estimate very precise RVs. This precision, combined with the modest level
of atmospheric variability expected in RSGs and the apparent lack of
short-period binary systems, makes RSGs ideal kinematic tracers of
their local populations.

In this article we study the RSG population of NGC\,330 using
multi-epoch spectroscopy with the goal of estimating the binary
fraction in the cluster. To do this, we employ spectroscopy from two
instruments where the data span baselines of more than a
year.  The RSG sample and observations are described in
Section~\ref{sec:obs}.  To estimate RVs for our targets, we employ a
novel slicing technique that specifically identifies the average
atmospheric velocity, making use of the huge number of spectral lines
available at optical wavelengths. Our method is outlined in
Section~\ref{sec:rv}, and building on previous
studies~\citep[][P19]{2016MNRAS.458.3968P}, we estimate the kinematic
properties and dynamical mass of NGC\,330.  We concentrate on the
multi-epoch measurements in Section~\ref{sec:rvv}, where we detail our
method of searching for significant variability for each target and
estimate the binary fraction of the sample.  In
Section~\ref{sec:discussion} we estimate the age of NGC\,330 based on
the distribution of RSG luminosities and summarise the key physical
properties of this cluster.  We present our main conclusions in
Section~\ref{sec:conclusion}.

\section{Observations}    \label{sec:obs}

Spectroscopic observations of RSGs in NGC\,330 are from a combination
of the Multi Unit Spectroscopic Explorer (MUSE; Bacon et al.  2014)
on the Very Large Telescope (VLT) and the High Accuracy Radial
velocity Planet Searcher \citep[HARPS,][]{HARPS} spectrograph on the
ESO 3.6m telescope.
The MUSE observations\footnote{Proposal ID: 60.A-9183(A)} cover the centre of the cluster, and their reduction is detailed in~\citet{2019arXiv191103477B}. Briefly, these data were reduced using the standard ESO MUSE pipeline v2.6.
A fraction of faintest pixels within the MUSE field of view is used to estimate the sky spectrum, which is subsequently subtracted from the data.
No telluric correction is necessary because the only prominent telluric features are within well-defined wavelength ranges that are taken into account when the RVs are estimated.
Source extraction is performed using Hubble Space Telescope (HST) photometry as an input catalogue, after first converting the HST photometry into the MUSE reference frame.
On this input catalogue, a tailored point-spread-function fitting routine is used to extract the flux of the sources at each wavelength step.

The MUSE observations span a baseline of over one year between 2017 August and
2018 December, with six separate epochs. Each epoch consists of five exposures of 540\,s, with 0\farcs7 offsets and a 90$^\circ$ derotator offset.
The observations were conducted in
the (wide-field) adaptive optics (AO) mode, providing spatial sampling
of 0\farcs2, a wavelength coverage of 4600-9300\AA, and a mean
spectral resolving power ($R$) of $\sim$3000. The typical
signal-to-noise ratios (S/N) of the RSG spectra are $\sim$25
around the calcium triplet ($\sim$8500\,\AA), which is the region
used to estimate RVs for the cool star spectra (see
Section~\ref{sec:rv}).

In addition to the recent MUSE observations, focused on the centre of
the cluster, we employ archival data from HARPS programmes that
targeted cool stars in the Magellanic Clouds\footnote{Proposal IDs:
  083.C-0413, 083.D-0549, 084.D-0591, 085.C-0614, 085.D-0395,
  086.D-0078.}.
The long-term stability of the HARPS spectrograph is known to be below 1\,$\mbox{m s}^{-1}$~\citep[e.g.][]{2012Natur.491..207D,2014Natur.513..358P} for the highest stability mode (the so-called EGGS mode has a precision of 3-5\,$\mbox{m s}^{-1}$ , according to the HARPS user manual).
The observations used the high-efficiency (wider
fibre) so-called EGGS mode, resulting in $R\sim$80\,000 over a
wavelength coverage of 3800--6900\,\AA.
Exposure times range between 1200 and 2400\,s depending on target and programme, which delivers a typical S/N of $\sim$5-15. 
The HARPS observations span a one-year
baseline between 2009 October and 2010 November, with a minimum of six
observations of each star enabling detection of short-scale variations
as well as longer-term trends. HARPS was designed to search
for extrasolar planets around cool low-mass stars \citep{HARPS}, and it therefore
delivers exquisite long-term RV stability and precision, which is
ideal for studying the binary fraction among RSGs.

\begin{figure}
  \centering
  \includegraphics[width=\columnwidth]{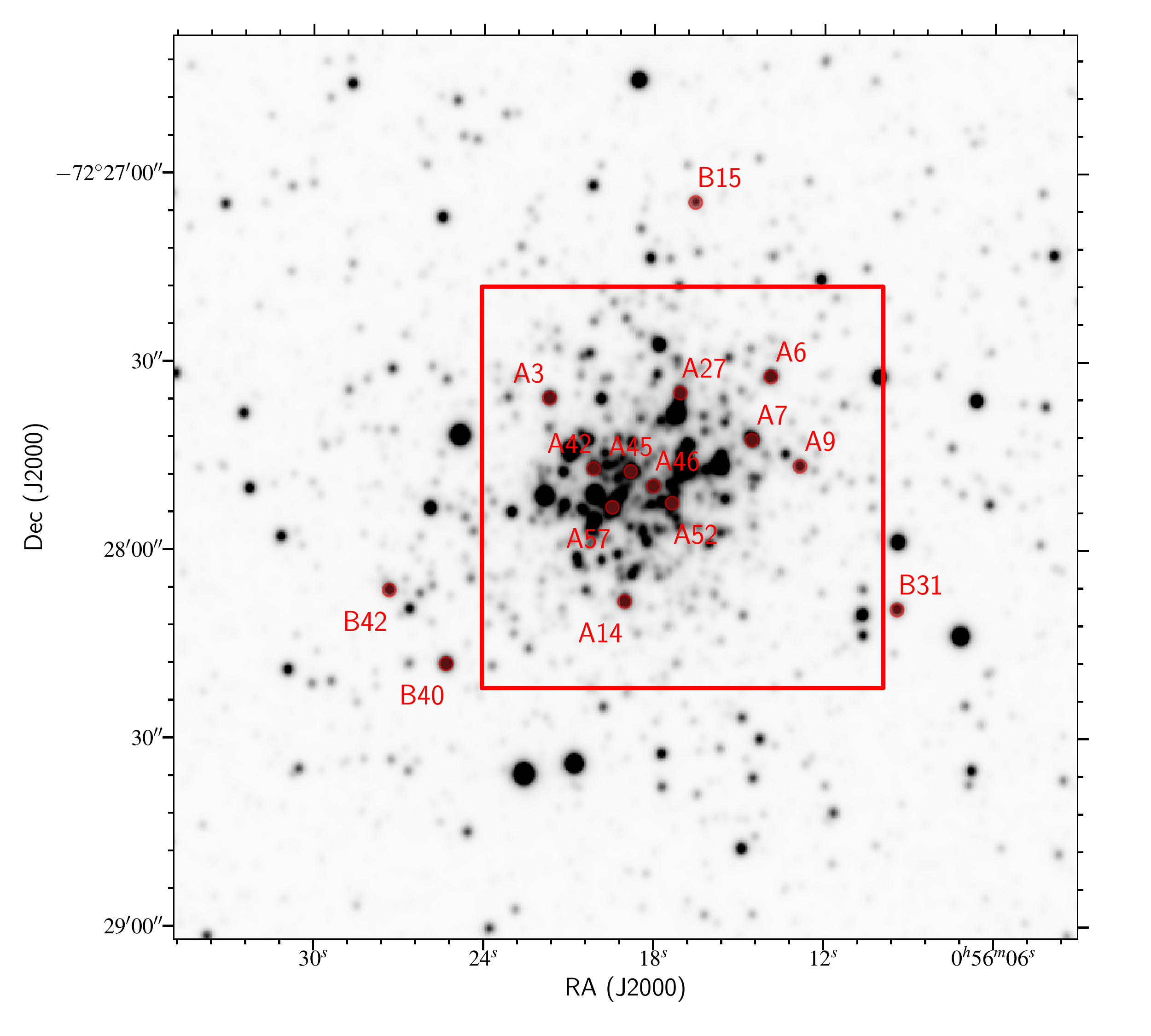}  
  \caption{$B$-band image of NGC\,330 obtained with the Wide Field Imager on
    the 2.2m MPG/ESO telescope \citep[see][]{eis01}. Red circles
    indicate our targets, with identifications primarily
    from~\cite{1974A&AS...15..261R}, see Table~\ref{tb:avRVs} for
    details. The approximate footprint of the observed MUSE field
    \citep{2019arXiv191103477B} is overlaid in red.}\label{fig:ngc330}
\end{figure}

The location of our targets is shown in Figure~\ref{fig:ngc330}. There are five
targets in common between the MUSE and HARPS data: A3, A6, A9,
A14, and A42 from \citet{1974A&AS...15..261R}. Given its relative
brightness and the interesting results for star B31 from the HARPS
data (see Section~\ref{sec:rvv}), we also extracted spectra from the wings
of the point spread function of this star from the edge of the reduced
MUSE datacube. The MUSE footprint also includes an additional six
RSGs: Rob74\,A7, A27, A45, A46, A52, and A57. The HARPS
observations add a further three RSGs: Rob74\,B15, B40, and B42.

\begin{figure}
  \centering
  \includegraphics[width=\columnwidth]{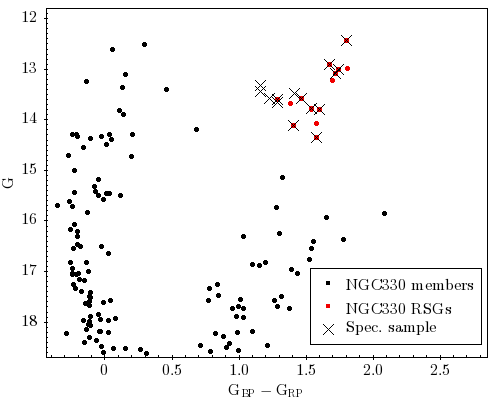}  
  \caption{{\em Gaia} $G_{BP}-G_{RP}$ vs. $G$ CMD of the field around NGC\,330. Black points highlight all targets that meet our NGC\,330 selection criteria, and red points highlight RSGs selected from the cluster members. Black crosses mark the RSGs with spectroscopic data presented here. This figure highlights the distinction between the RSG populations and lower mass red giant stars as well as blue supergiants in the cluster. Three stars in the spectroscopic sample do not meet the {\em Gaia} selection criteria; see text for details.}\label{fig:cmd}
\end{figure}

In total, we present spectroscopic observations for 15 RSGs in
NGC\,330. Spectral types for our targets were determined from the
HARPS and MUSE spectra following the classification criteria and method detailed in~\citet{2018A&A...618A.137D},
with the average spectral types for each star listed in Table~\ref{tb:avRVs}.
All targets were cross-matched with {\em Gaia} DR2 \citep[Gaia
collaboration,][]{2018A&A...616A...2L} to assess membership based on
proper motions and parallaxes. The {\em Gaia} colour-magnitude diagram (CMD) of the cluster is shown in Figure~\ref{fig:cmd}, where the potential RSGs in NGC\,330 are highlighted in red and the 15 RSGs considered here are also highlighted.

To construct this CMD, we extracted all {\em Gaia} sources within 2\farcm5 from the cluster centre (roughly corresponding to the size of Figure~\ref{fig:ngc330}) and cleaned the sample for parallax and proper motion to select SMC-like candidates.
In general, the different populations present in this diagram are in good agreement with those from \cite{1985AJ.....90.1196C}. We note that a clear gap exists between the RSG population and that of the blue supergiant population at $(B - R)$\,$=$\,~0-0.2 and $G$\,$>$\,15, as well as between the RSGs and a population of likely field stars at $(B-R)$\,$>$\,0.4 and $G$\,$>$\,16~\citep{1985AJ.....90.1196C}.

All 15 targets in the spectroscopic sample of RSGs have proper
motion measurements that are consistent with the bulk movement of the
SMC~\citep{2018A&A...616A..12G}.
Of the 15 targets, 11 have parallax measurements consistent with zero at the two-sigma level.
Four targets (A42, A45, A52, and A57) have significant parallax measurements, but the uncertainties are larger than the average. This indicates that these targets are potential Galactic contaminants.
However, these four targets are located very near the cluster centre, where potential contamination is more likely as a result of crowding. 
One target (A46) has a right ascension proper motion measurement that is formally outside our {\em Gaia} NGC\,330 membership criteria. It is therefore not included as an RSG member in Figure~\ref{fig:cmd}, but we consider this target a genuine member given that the one-sigma uncertainties on the measurement take it within our proper motion threshold. 
Based on these data, we conclude that
at least 11 of 15 of the targets are genuine SMC members, that is, those with
reliable kinematic information from {\em Gaia}.
The four targets with less reliable {\em Gaia} data (A42, A45, A52, and A57) are indicated in Table~\ref{tb:avRVs} as potential Galactic contaminants and are included in the sample for further study. 

\begin{table*}
\caption{Observational information and mean radial velocities ($\overline{v}$) for RSGs in NGC\,330. Targets selected by our {\em Gaia} criteria without HARPS or MUSE spectra are listed in the No spectra section.}
\label{tb:avRVs}      
\centering                                      
\begin{tabular}{lccccclc}          
\hline\hline                        
ID & \multicolumn{2}{c}{ number of epochs} & $J$ & $H$ & $K$ & SpT & $\overline{v}$\,$\pm$\,$\sigma$\\    
& HARPS & MUSE &  & & & & (\kms)  \\
\hline
\multicolumn{8}{c}{Spectroscopic sample} \\
\hline

A3      &  16 & 6  & 12.019 &  11.469 & 11.374  &  G3\,Ib     & 151.5\,$\pm$\,0.5 \\
A6      &  10 & 6  & 10.969 &  10.217 & 10.087  &  K1\,Iab    & 155.3\,$\pm$\,1.1 \\
A7      & \z0 & 6  & 10.193 & \z9.408 & \z9.182 &  G6 Ib      & 152.4\,$\pm$\,0.9 \\
A9      & \z9 & 6  & 11.778 &  11.072 & 10.874  &  K1\,Ib     & 154.4\,$\pm$\,0.9 \\
A14     & \z9 & 6  & 10.908 &  10.15\z& \z9.932 &  K1\,Ib     & 154.6\,$\pm$\,1.9 \\
A27     & \z0 & 6  & 11.509 &  10.636 & 10.474  &  K1\,Ib     & 149.1\,$\pm$\,2.4 \\
A42$^*$ & \z6 & 6  & 11.76\z&  11.03\z& 10.855  &  K0\,Ib-II  & 153.9\,$\pm$\,0.6 \\
A45$^*$ & \z0 & 6  & 11.592 &  10.916 & 10.719  & G3.5 Ia-Iab & 152.0\,$\pm$\,2.2 \\
A46     & \z0 & 6  & 11.643 &  10.952 & 10.653  &  K0\,Ib     & 157.5\,$\pm$\,1.4 \\
A52$^*$ & \z0 & 6  & 11.517 &  11.009 & 10.745  &  K0\,Ib     & 162.1\,$\pm$\,4.0 \\
A57$^*$ & \z0 & 6  & 11.296 &  10.660 & 10.412  &  G7 Ia-Iab  & 154.6\,$\pm$\,1.5 \\
B15     & \z7 & 0  & 12.38\z&  11.64\z& 11.518  &  K0\,Ib     & 149.8\,$\pm$\,0.6 \\
B31     & \z7 & 1  & 12.233 &  11.491 & 11.339  &  K\,II      & \z107.8\,$\pm$\,14.6 \\
B40     & \z7 & 0  & 10.92\z&  10.199 & 10.032  &  G7\,Iab-Ib & 153.1\,$\pm$\,1.1 \\
B42     & \z8 & 0  & 11.842 &  11.158 & 10.991  &  G8\,Ib-II  & 151.9\,$\pm$\,0.5 \\
\\
\multicolumn{8}{c}{No spectra} \\
\hline
B10         & \z0 & 0 & 10.784 & 10.024 &   9.864  & G8.5\,Ia-Iab & 148.0\,$\pm$4.0\\
B19         & \z0 & 0 & 11.155 & 10.437 &  10.291  &    --           & --  \\
B20         & \z0 & 0 & 11.995 & 11.421 &  11.316  &    --    & --   \\
ARP III-210 & \z0 & 0 & 11.934 & 11.226 &  11.084  &    --           & --  \\
ARP III-214 & \z0 & 0 & 12.135 & 11.440 &  11.276  &    --           & --  \\

\hline                                             
\end{tabular}
\tablefoot{Literature identification from \citet[][Rob74]{1974A&AS...15..261R} and \citet[][ARP]{1959AJ.....64..254A}.
See Table~\ref{tb:IDs} for more information on identification.
An asterisk$^{}$~identifies a potential Galactic contaminant based on {\em Gaia} data.
Spectral type and RV data for B10 are from~\citet{2015A&A...578A...3G} because no HARPS or MUSE spectra are available.}
\end{table*}

\section{Radial velocities}    \label{sec:rv}

The RVs for our targets were estimated using a similar iterative
cross-correlation approach as was used by P19.
This approach splits the spectra into small wavelength slices that are assumed to provide an independent estimate of the average RV of the star.
The slices were compared to a synthetic spectrum extracted from
a {\sc marcs} stellar model atmosphere from the non-local thermodynamic equilibrium database hosted
by the Max Planck Institute for Astronomy\footnote{http://nlte.mpia.de/}.
The distribution of RVs estimated from these slices was used to define the RV of the star as well as the uncertainties on this measurement, where a rigorous sigma-clipping
approach was used to remove any outlying slices.

To estimate RVs, we used the 6000$-$6800\,\AA~ range for the HARPS spectra and the 6200$-$8800\,\AA\ range for the MUSE spectra. These are the reddest parts of the spectra in both cases because this is where the S/N is highest.
Within these wavelength regions the spectra were split into
42\,\AA~and 167\,\AA~ wide wavelength ranges for the HARPS and MUSE spectra, respectively.
The synthetic spectrum used to compare to the observation was extracted from an appropriate {\sc marcs} model with stellar parameters for an SMC-like RSG
(i.e. T$_{\rm eff}$~=~4000\,K, $\log$g~=~0.0\,dex, [Z]~=~$-$0.75\,dex), and
the parameters were convolved by the relevant broadening function for comparison
with the observed spectral resolution of the HARPS and MUSE data.
The RV for the star is taken to be the mean of the clipped distribution of RVs from the slices.
The uncertainty on this measurement is the standard error on the mean ($\sigma/\sqrt{N_{\rm f}}$, where
$N_{\rm f}$ is the number of retained slices).
The typical uncertainties on the HARPS and MUSE data is 0.09 and 1.1\,\kms, respectively.

We estimated RVs with the method outlined above because various intrinsic effects perturb the RVs of cool stars as a result of stellar activity in addition to the instrumental effects that can introduce systematic offsets. 
One of the main sources of uncertainty in estimating RVs for these targets is atmospheric variations.
Given the significant atmospheric extension of RSGs, groups of atomic absorption
features that are formed at similar depths
can display significantly different velocities, with relative differences between groups as large as 25\,\kms~\citep[][although the effect of this on the overall RV estimate is significantly lower]{2007A&A...469..671J,2019arXiv191004657K}.
Estimating RVs using only a small number of spectral features can be biased towards atmospheric motions, particularly if these features are dominated by a group of lines at a particular depth.

\begin{figure}
  \centering
  \includegraphics[width=\columnwidth]{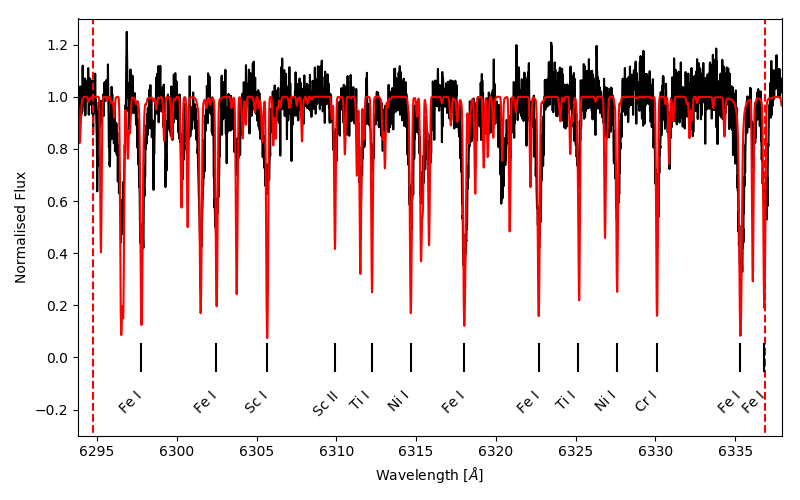}  
  \caption[]{\label{fig:slices} %
    Example slice used in the RV estimation for the HARPS data.
    Some of the strongest lines are highlighted, demonstrating
    the density of spectral features present in a typical slice.  The
    template spectrum calculated from a {\sc marcs} model atmosphere
    is overlaid in red, and the red dashed vertical lines highlight the edges
    of the slice.}
\end{figure}

To attempt to mitigate these effects, we specifically selected the slice width to include a significant number of atomic absorption features.
Figure~\ref{fig:slices} illustrates the absorption features present in a typical slice.
A full analysis of the depths at which these features form is beyond the scope of the current study, but we assume that the RV estimate from each slice tends towards the average RV of the star and not the atmospheric variation at a given depth.
This is supported by the fact that our sigma-clipping routines remove between 0 and 5 slice measurements from each of the final calculations, compared with a total number of 20 slices in the HARPS spectra. Larger variations and accordingly more clipped slices would be expected if atmospheric variability significantly perturbed many slices. 
In addition, we see no evidence for a wavelength dependence in the RVs estimated from individual slices, but we note that undetected atmospheric variability is the dominant source of uncertainty in the quoted RV measurements on individual epochs.

We also evaluated our choice of synthetic spectrum.
To test the extend of the effect of choosing a synthetic spectrum on the precision and stability of the RV estimates, we repeated the analysis using a synthetic spectrum that was extracted with a different radiative transfer code~\citep[{\sc turbospectrum};][]{2012ascl.soft05004P}, with a different line list and
stellar parameters.
The results of this analysis are that there are no significant differences in the average RVs for each target. 
These comparisons reveal a systematic offset of $\sim$0.35\kms\  in the RV estimates of the HARPS data using different synthetic spectra; this difference is undetectable in the MUSE data.
This systematic does not affect the internal consistency of the RV estimates, and we therefore conclude that our model is an appropriate choice for these observations, but that on average, our absolute RV cannot be considered more accurate than $\pm$0.35\,\kms.
We find that in practice, RV variations that we attribute to atmospheric variation dominate the dispersion of our multi-epoch data, and Table~\ref{tb:avRVs} shows that no targets have a standard deviation smaller than 0.5\,\kms. However, we note that 11 out of our 15 targets have a dispersion smaller than 2\,\kms, which suggests that atmospheric variability has a limited effect on our measurements.

The accuracy of our uncertainties on individual measurements was tested by comparing the results using another well-tested method to robustly estimate uncertainties~\citep[e.g.][]{2007AJ....134.1843A}.
To do this, we fitted the peak of the cross-correlation function of the observed spectrum with the model spectrum using a combination of a Gaussian function and a low-order polynomial. The best-fitting model was estimated using a Levenberg-Marquardt algorithm and the least-squares statistic.

When the cross-correlation function of large regions of the spectrum is fitted in the same wavelength regime as was used by the slice analysis, the estimated uncertainties compare reasonably well with those estimated using the slice technique. Typically, the uncertainties estimated using the RV distribution from the slices are a factor of 1 --- 2 smaller than the uncertainties estimated by fitting the peak of the cross-correlation function. This adds strength to our uncertainty estimates on individual epochs. However, as noted above, inaccuracies in the synthetic spectrum used and intrinsic atmospheric motions dominate the estimated uncertainties.

To ensure there are no systematic offsets between the MUSE and HARPS results, in the MUSE data the strong telluric absorption
features were used to provide an absolute calibration to the RVs,
similar to methods used at longer
wavelengths~\citep[e.g.][]{2015ApJ...798...23L,2015ApJ...803...14P,2017MNRAS.468..492P}.
To do this, a model telluric spectrum was generated using the {\sc molecfit}
tool~\citep{2015A&A...576A..77S}.
After this correction, we find no significant differences between the HARPS and MUSE spectra.
In this analysis, we consider the uncertainties from instrumental stability of the HARPS spectra as negligible in comparison to that of MUSE.

Literature measurements exist for about 50\% of our targets.
In general, our results compare well with previous measurements~\citep{1980MNRAS.191..285F,1985AJ.....90.1196C,1991A&A...252..557S,1999AJ....117.2286G,1999A&A...345..430H,2015A&A...578A...3G}.
See Figure~\ref{fig:rv_lit} for a comparison.
These comparisons and those in P19 provide us with confidence that this technique provides accurate and precise RV measurements and uncertainties.

\subsection{Kinematic analysis}    \label{sub:kinematics}

\begin{figure}[btp]
  \centering
  \includegraphics[width=\columnwidth]{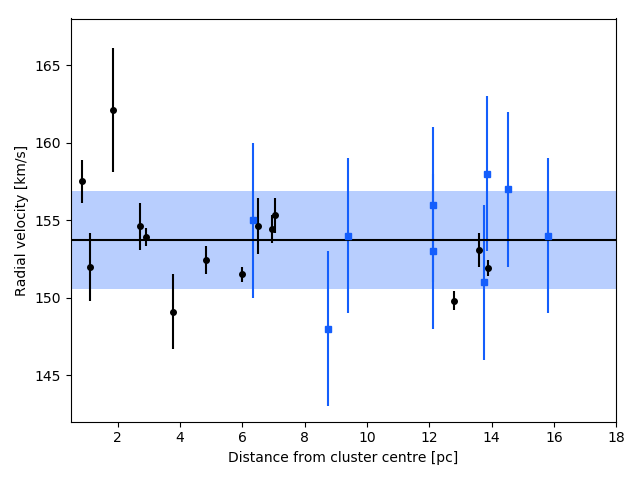}  
  \caption[]{\label{fig:rvsRV} %
  Average radial velocities as a function of distance from the centre of the cluster \citep[as defined by][]{2003MNRAS.338...85M}. 
  Black points show average velocities of the RSGs with RV estimates from Table~\ref{tb:avRVs}, and blue squares show OBA-type stars from~\citet{2006A&A...456..623E}.
  The black solid line and blue shaded region illustrate the systemic velocity and velocity dispersion of the cluster estimated using a combination of RSGs and OBA-type stars (153.7\,$\pm$3.2\,\kms; where the uncertainty quoted and displayed here is the $\sigma_{1D}$ measured in section~\ref{sub:kinematics}). The black points with the largest uncertainties are those with only MUSE data.
  }
\end{figure}

Average RVs for each target are listed in Table~\ref{tb:avRVs}, where
the quoted uncertainties are unbiased estimates of the standard deviation
from the multi-epoch observations, taking into account the sample size and measurement errors.
Figure~\ref{fig:rvsRV} displays these results
(excluding B31, see Section~\ref{sec:rvv}) as a function of projected distance
from the centre of the cluster~\citep[as defined
by][as $\alpha$\,$=$\,00:56:18.0, $\delta$\,$=$\,$-$72:27:47.0, J2000]{2003MNRAS.338...85M}.
The RVs of all targets are consistent with cluster membership.
Additional analysis based on the
probability of membership of either NGC\,330 or the SMC field was
attempted, but given the similarities and overlap between the
velocities of these structures, a meaningful comparison was not
possible.

Following~\cite{2016MNRAS.458.3968P}, we estimated the line-of-sight
velocity and velocity dispersion of NGC\,330 using {\sc
  emcee}~\citep{2013PASP..125..306F}, which is an implementation of
the ensembler sampler for the Markov chain Monte Carlo method
of~\cite{2010CAMCS.5..65G}.  The likelihood function is identical to
that of~\cite{2016MNRAS.458.3968P}, where the implicit assumption is
that the intrinsic velocity dispersion of the cluster is Gaussian in
nature and constant over the radial distance covered by our targets.

After excluding all obviously variable sources (in this case, only B31 was excluded),
our estimated line-of-sight velocity for NGC\,330 (v$_0$) is
(153.7\,$\pm$1.0\,\kms), with a line-of-sight velocity dispersion
($\sigma_{1D}$) of 3.46\,$^{+0.88}_{-0.61}$\,\kms.
When we exclude the four potential Galactic contaminants, v$_0$ and $\sigma_{1D}$ do not change significantly.

NGC\,330 also hosts a significant population of B-type stars; RV
estimates for seven stars were given by \cite{2003A&A...398..455L}.  RVs
for a larger sample of $\sim$100 early-type stars in the outskirts and
field around the cluster were given by \citet{2006A&A...456..623E}.
Figure~\ref{fig:rvsRV} also includes the RVs for the apparently single
OBA-type stars from \cite{2006A&A...456..623E} that lie within 18\,pc
from the cluster centre. Notwithstanding the larger uncertainties
on the values for the hotter stars, there is generally good agreement between the two samples.

\cite{1980MNRAS.191..285F} estimated $\sigma_{1D}$ for NGC\,330 using
both early- and late-type stars, but their results were
limited by the precision in their RV measurements. We adopted a similar approach of
combining our cool-star RVs with those for early-type stars from
\citeauthor{2006A&A...456..623E} Although they potentially trace
slightly different populations~\citep{2019arXiv191103477B} and have
larger uncertainties, the OBA stars have similar systemic velocities
as the cool stars, and we used them to bolster our sample for the
kinematic analysis.
After excluding obviously variable OBA-type stars from the sample, we estimate
v$_0$~=~153.7\,$\pm$\,0.8\,\kms\ and
$\sigma_{1D}$\,$=$\,3.20\,$^{+0.69}_{-0.52}$\,\kms\ (in good agreement
with the estimate from the RSGs alone).
Again, excluding the four potential Galactic contaminants, v$_0$ remains unchanged, whereas $\sigma_{1D}$ is slightly reduced (2.54\,$^{+0.63}_{-0.48}$\,\kms).

Given their comparative lack of RV
variability~\citep[e.g.][P19]{2007A&A...469..671J}, RSGs are important
tracers of the kinematic properties of external young massive
clusters. B-type stars~\citep[which constitute the majority of the
sample from][]{2006A&A...456..623E} display intrinsic RV variations
within their atmospheres of up to approximately
15\,\kms~\citep{2014MNRAS.442.1483T,2015A&A...580A..93D}.
Despite this and the
significantly larger uncertainties on their RVs, including them slightly decreases the uncertainties on
$\sigma_{1D}$.  This means that including these targets increases
the reliability of the $\sigma_{1D}$ measurement because the
uncertainty on the dispersion is dominated by the number of objects.

\subsection{NGC330 mass estimates} 
\label{sub:mass}

When we assume the virial theorem, the determination of $\sigma_{1D}$ allows the dynamical mass (M$_{\rm dyn}$) of the cluster to be estimated, as follows:

\begin{equation}
  {\rm M_{dyn}} = \frac{\eta\sigma_{1D}^{2}r_{\rm eff}}{G}~,
  \label{eq:vir}
\end{equation}

\noindent where $r_{\rm eff}$ is the effective radius and $\eta$~=~$6r_{\rm
  vir}/r_{\rm eff}$~=~11.0, using the definitions for $r_{\rm vir}$
and $r_{\rm eff}$ from~\citet[][and references
therein]{2010ARA&A..48..431P} of 11.14 and 6.11\,pc, respectively.
The dynamical mass estimated for NGC\,330 using the velocity
dispersion of  RSGs alone is
log\,(M$_{\rm  dyn}$/M$_{\odot}$)~$=$~5.27\,$\pm$\,0.20.
A previous estimate of
the dynamical mass of NGC\,330 is listed by~\cite{2010ARA&A..48..431P}
as log\,(M$_{\rm dyn}$/\,M$_{\odot}$)~=~5.64, based on the King profile fits of  \cite{2005ApJS..161..304M}.

As shown by P19, RSGs are robust tracers of the dynamical properties
of clusters as a result of their small intrinsic RV variations.   In light of the limitations
based on the number of RSGs within the cluster, we also estimated the
dynamical mass using the OBA-type stars
from~\citet{2006A&A...456..623E} within 18\,pc of the cluster centre to add
strength to the dynamical mass estimate.
By combining the two samples, we obtain a dynamical mass estimate of
log\,(M$_{\rm dyn}$/M$_{\odot}$)\,$=$\,5.20\,$\pm$\,0.17, again in
good agreement with the estimate from RSGs alone.
This is significantly larger than the photometric mass estimate of $\log
(M_{phot}/M_{\odot}$)~$=$~4.58\,$\pm$\,0.2 from
\citet{2003MNRAS.338...85M} and that of~\citet{2019arXiv191103477B}.
This is to some extent expected because M$_{\rm dyn}$ measurements made in this way are in general affected by binarity. 
\citet{2010MNRAS.402.1750G} showed that they were able to reproduce the difference in dynamical mass and photometric mass estimates for young clusters such as NGC\,330 with a binary fraction among the supergiant population of 25\%, which is in good agreement with our binary fraction estimate (presented in Section~\ref{sec:rvv}).

\section{RV variability and multiplicity analysis} 
\label{sec:rvv}

Long-baseline RV variability studies have been conducted for only a
handful of individual Galactic RSGs, for example $\alpha$~Ori
\citep{1989AJ.....98.2233S}.
\cite{2007A&A...469..671J} studied well-known Galactic RSGs with the goal of identifying
atmospheric velocity variability. By calculating RVs of their
sample, these authors identified two groups of variability that were split at
$\sim$5\,\kms, where $\sim$50\% of their sample displayed RV
variability above this limit.

P19 studied the RV variation of a sample of 17 RSGs within the
30~Doradus region as part of the VLT-FLAMES Tarantula
Survey~\citep[VFTS; ][]{2011A&A...530A.108E}. The time sampling of the
spectroscopy was comparable to our data in NGC\,330, and to estimate
the RVs, they developed the technique used here (with greater precision
possible here due to the high resolution of HARPS). The results from
P19 are somewhat at odds with the study of \citet{2007A&A...469..671J}
in the sense that only 1 of 17 sources displayed a RV variation higher than 5\,\kms.

The expected level of RV variation for a single RSG as a result of the
convective motions within their atmospheres is 1--5\,\kms\
\citep{1975ApJ...195..137S}, which has been demonstrated
observationally in multiple
studies~\citep[P19]{1928MNRAS..88..660S,1933ApJ....77..110S,1989AJ.....98.2233S,2007A&A...469..671J,2010ApJ...725.1170S}.

The expected level of RV variation for an RSG within a binary system is
more difficult to predict.
This is because there
are relatively few literature examples of genuine RSGs within binary
systems~\citep[e.g.][and fewer still with accurate RV
measurements]{2017ars..book.....L,2019ApJ...875..124N}. P19
estimated the expected semi-amplitude variation ($K$) in the RV for an RSG
primary with masses of 8 and 15\,M$_\odot$ and found that $K$ must be
between $\sim$2--30\,\kms\ with orbital periods greater than 2.5\,yr
for an 8\,M$_\odot$ RSG, rising to 4.5\,yr for 15\,M$_\odot$.

\begin{figure*}[hbtp]
  \centering
  \includegraphics[width=\linewidth]{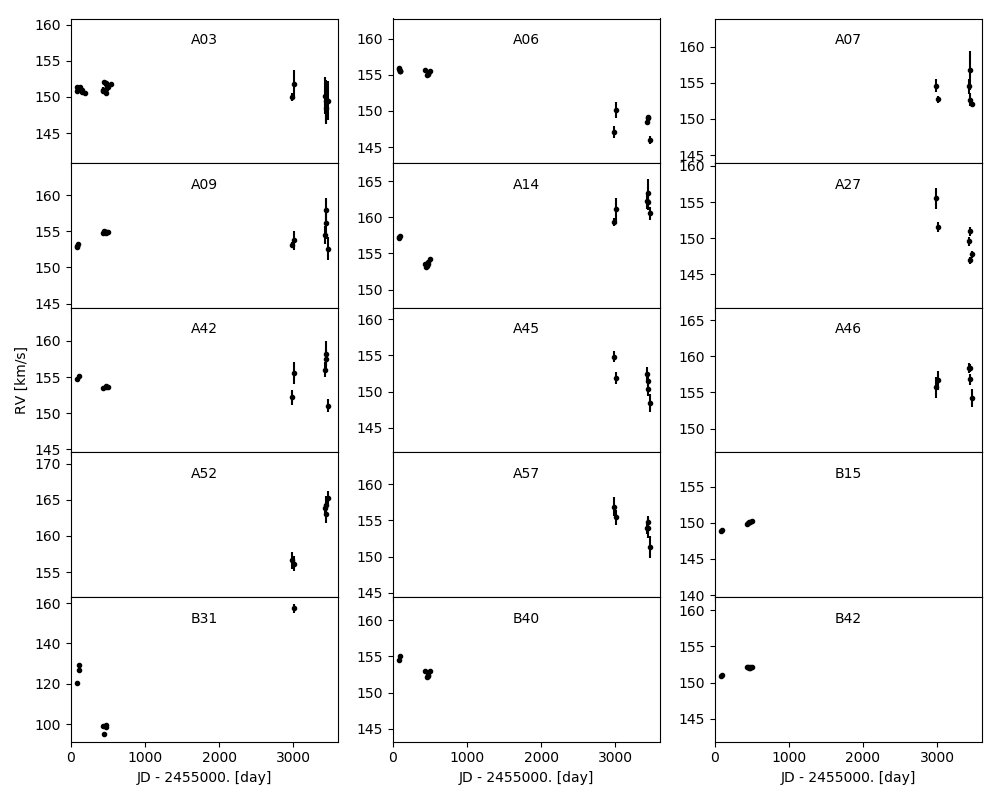}  
  \caption[]{\label{fig:rv_all} %
  RV estimates for all targets, shown on the same relative scale to demonstrate the observed differences in RV variations (except for Rob74~B31, which displays significant RV variability; see text for details).
}
\end{figure*}

Figure~\ref{fig:rv_all} shows the RV
estimates of our targets from the HARPS and MUSE spectra.   Except for B31, all panels show the same relative abscissa.
Table~\ref{tb:rvall} lists the RV estimates for all epochs of each
target.
The most precise RV measurements are achieved from the HARPS data where the uncertainties reach as
low as $\sim$60\,m\,s$^{-1}$, where the precision is, in general, determined by the S/N of the spectra.

B31 is a known binary system~\citep{2007A&A...472..577M} and will be
discussed in more detail by Patrick et al. (in prep).  Except for B31, all targets display indications of some
low-level variation, with the maximum difference between two epochs in
any given target of 4.1\,\kms\ (A14), in the HARPS data.
The general variability in our
sample is in good agreement with the levels seen in one half of the sample
of~\cite{2007A&A...469..671J}, as well as in the highest quality data of
P19.

A3 is the best sampled target and displays
RV variations with an amplitude of $\sim$1\,\kms in the HARPS data, and with larger-than-average uncertainties, perhaps connected to the fact that it has the
earliest spectral type of the sample: it is classified as G3~Ib.
There is no detectable trend of RV uncertainty as a function of spectral type.
A14, B40,
and A42 potentially have signatures of periodic variation in their RV
estimates. B15 and B42 have continually increasing RV values over the
course of the observational campaign, similar to the trends observed
in two stars of~\cite{2007A&A...469..671J}.  Unfortunately, neither
target has follow-up MUSE observations that might constrain their
longer term variability.

In the MUSE data, the uncertainties are significantly larger.
However, the comparison between the average RVs for targets with both
MUSE and HARPS data is generally quite good.  For three of the six
targets in common, the difference between average RVs is $\sim$
1.0\,\kms.  The three stars outside this range are A3, A14, and B31,
all of which are potentially variable. We therefore conclude that the
agreement between the HARPS and MUSE data is within the MUSE
uncertainties.

\subsection{Multiplicity analysis}
\label{sub:varb}

It is important to distinguish, if possible, between RV variability from
atmospheric motions (convection, pulsations) and that of
variability arising from binary motions. In the case
of RSGs, this is difficult to achieve because of the types of binary
systems that are expected~\citep{1977JRASC..71..152W} and because of the expected level of
variability from their
atmospheres~\citep{1928MNRAS..88..660S,1989AJ.....98.2233S,2007A&A...469..671J,2010ApJ...725.1170S,2010A&A...515A..12C,2011A&A...528A.120C}.

Previous studies that aimed to detect variability owing to binarity where atmospheric variability is a contaminating factor~\citep[e.g.][P19, Lohr et al. in prep.,
Patrick et al. in prep.]{2013A&A...550A.107S,2015A&A...580A..93D} used two criteria
to detect significant RV variations:

\begin{equation}
\label{eq:var}
\frac{|v_i - v_j|}{\sqrt[]{\sigma_i^2 + \sigma_j^2}} > 3.0, $~and~$ |v_i - v_j| > \Delta RV_{min}~,
\end{equation}

where $v_i$, $\sigma_i$ ($v_j$, $\sigma_j$) are the measured RVs and
corresponding uncertainty at epoch $i$ ($j$) at the epochs ($i$ and $j$) providing the maximum velocity difference.  The
first criterion searches for significant variability at the 3$\sigma$
level, taking into account the uncertainties on the measurements.
The second criterion ($\Delta RV_{min}$) places a limit on the
difference between any two given measurements.

\begin{figure*}[btp]
  \centering
  \includegraphics[width=0.6\textwidth]{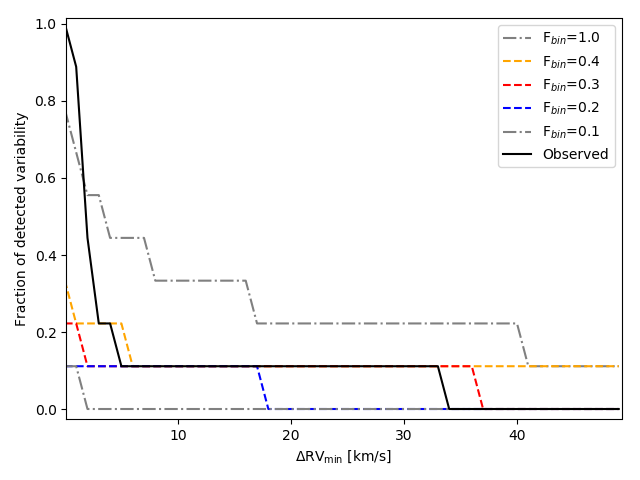}  
  \caption[]{\label{fig:bf} %
    Observed fraction of detected variability of the RSGs in NGC\,330 as a function of
    the $\Delta$RV$_{min}$ parameter for the nine RSGs with HARPS data (black solid line).
    The three best-fitting simulated stellar populations with an empirically defined
    distribution from~\citet{2017ApJS..230...15M}, containing between
    20\% and 40\% binary systems with orbital periods in the range
    2.3~$<\log P [{\rm day}] <$~4.3, are shown with dashed coloured
    lines.
    The dot-dashed grey lines show the limits of simulations containing a binary fraction of 10\% and 100\%. The time sampling, uncertainties, and number of targets of the simulations is chosen to match the HARPS data. Each simulated curve is an average of 10,000 iterations, but to best compare with the observations, the simulated curves are discretised to the same intervals as the observations.}
\end{figure*}

However, given the precision of the HARPS measurements combined with the observed variability as a result of atmospheric variations, the first criterion is always met for these data. Therefore, our variability analysis is based solely on the second variability criterion above. This criterion essentially searches for the largest difference between measured RV measurements, which is an appropriate and effective method for detecting variability, particularly in the regime of sparsely sampled RV curves~\citep[e.g.][]{2012ApJ...751..143M,2013A&A...550A.107S,2018ApJ...854..147B}.
Figure~\ref{fig:bf} illustrates the variation in the number of targets meeting the
selection criteria as a function of the $\Delta RV_{min}$
parameter for HARPS data.

Rather than select a specific value of $\Delta RV_{min}$ to
distinguish between variations from atmospheric motions of single
stars and variability from binarity, we chose to evaluate the binary
fraction of RSGs in NGC\,330 by simulating stellar populations with a
range of input binary fractions. The binary simulations used
an empirically defined range of parameters
from~\citet{2017ApJS..230...15M} for a flat distribution of orbital
periods between 2.3\,$<$\,log\,P\,[days]\,$<$\,4.3. The mass-ratio
probability density function was defined using a broken power-law
distribution defined as $p_q \propto q^{-\gamma}$, where
$\gamma$~=~$-$1.7 for $q >$~0.3 and $\gamma$~=~$-$0.2 for 0.1~$< q
<$~0.3.  
Primary masses were drawn from a Salpeter initial mass function, over the range  7\,$<$\,M$_1$/\,M$_\odot$\,$<$\,25, which was chosen to roughly correspond to reproduce the range in masses observed in NGC\,330.
The eccentricities of the systems were defined based on a
power-law distribution probability density function with an
exponent of +0.8 in the range 0.0~$< e <$~0.8, and the inclination of the system was drawn at random from angles between 0 and 90\,$^{\circ}$.

Using these parameter distributions, we drew a sample of binary systems with the time sampling, uncertainties, and sample size of the HARPS observations.
By solving the positions of each system over time, we estimated the RVs for each epoch, and using the uncertainties, these simulated observations were tested for significant variability using the same criterion as the observations.
This process was repeated 10\,000 times for populations with intrinsic binary fractions in the range from 10\%\ to 100\%.
These results are shown alongside the observed distribution in Figure~\ref{fig:bf}, where the shapes of the distributions reflect the sample size.

At small $\Delta RV_{min}$, the
number of significant detections increases to almost 100\%.  This is a
result of the small-scale variations seen in the data and is as
expected for a typical RSG.  At small $\Delta RV_{min}$, rather than probing
the observed binary fraction, the analysis would therefore be contaminated
heavily by atmospheric variations of single RSGs. As $\Delta RV_{min}$
increases, the number of stars displaying significant variability
decreases sharply.
By comparing the dashed line in Figure~\ref{fig:bf}, which corresponds to simulations with a binary fraction of 100\%, we see that this curve still does not reproduce the abundance of detected variability at small $\Delta RV_{min}$, nor the sharp drop at around $\Delta RV_{min}$ =~2--4\,\kms.
This is further evidence that the variability on this scale is a result of atmospheric velocities arising from single RSGs, in agreement with observations.

To evaluate the intrinsic binary fractions that best reproduce the observed distribution of detected systems as a function of the $\Delta RV_{min}$ parameter, we ignored the measurements below $\Delta RV_{min}$~=~4\,\kms\ because below this value, we assume that atmospheric variations dominate. The goodness of fit was assessed using a $\chi$-squared technique. 
By averaging the three best-fitting models, we estimate a binary fraction for the RSG population of NGC\,300 at $f_{\rm RSG}$ = 0.3\,$\pm$\,0.1, within the range 2.3\,$<$\,log\,P\,[days]\,$<$\,4.3, $q >$~0.1.
All binary fractions greater than 30\,\% over-predict the number of systems detected at large $\Delta RV_{min}$, where we expect that our observations are reasonably complete.

This estimate is in good agreement with the estimate from P19 in the
Tarantula region of the LMC.  However, P19 considered a smaller range
of periods (3.3\,$<$\,log\,P\,[days]\,$<$\,4.3) and mass ratios ($q >$~0.3).
P19 argued that the size
of RSGs excludes orbital periods shorter than
log\,P\,[days]\,$=$\,2.9, assuming a radius appropriate for an LMC-like
8\,M$_\odot$ RSG~\citep{2011A&A...530A.115B}.  It is well documented
in the literature that the average spectral type of RSGs is dependent
upon
metallicity~\citep{1985ApJS...57...91E,MasseyOlsen03,2012AJ....144....2L,2018MNRAS.476.3106T},
and given the masses implied by the spread of luminosities among the
RSG population of NGC\,330 (see Section~\ref{sub:evo}), we therefore
considered orbital periods down to $\sim$200\,d.  This is supported by
the orbital period reported for B31 of $\sim$455\,d~\citep{2007A&A...472..577M}. The binary fraction of RSGs in
NGC\,330 does not depend strongly on the range of orbital periods
considered. For example, when we extend our analysis to orbital periods
in the range 2.3\,$<$\,log\,P\,[days]\,$<$\,5.3, we find $f_{\rm RSG}$ =
0.30\,$\pm$\,0.10.
\citet{2017ApJS..230...15M} estimated a binary fraction of $\sim$0.3 for O- and B-type primaries over this range of orbital periods, again in good agreement with the estimate presented here.


\section{Discussion}    \label{sec:discussion}
\subsection{RSG age of NGC\,330}
\label{sub:ngc330properties}

\begin{figure}[btp]
  \centering
  \includegraphics[width=0.5\textwidth]{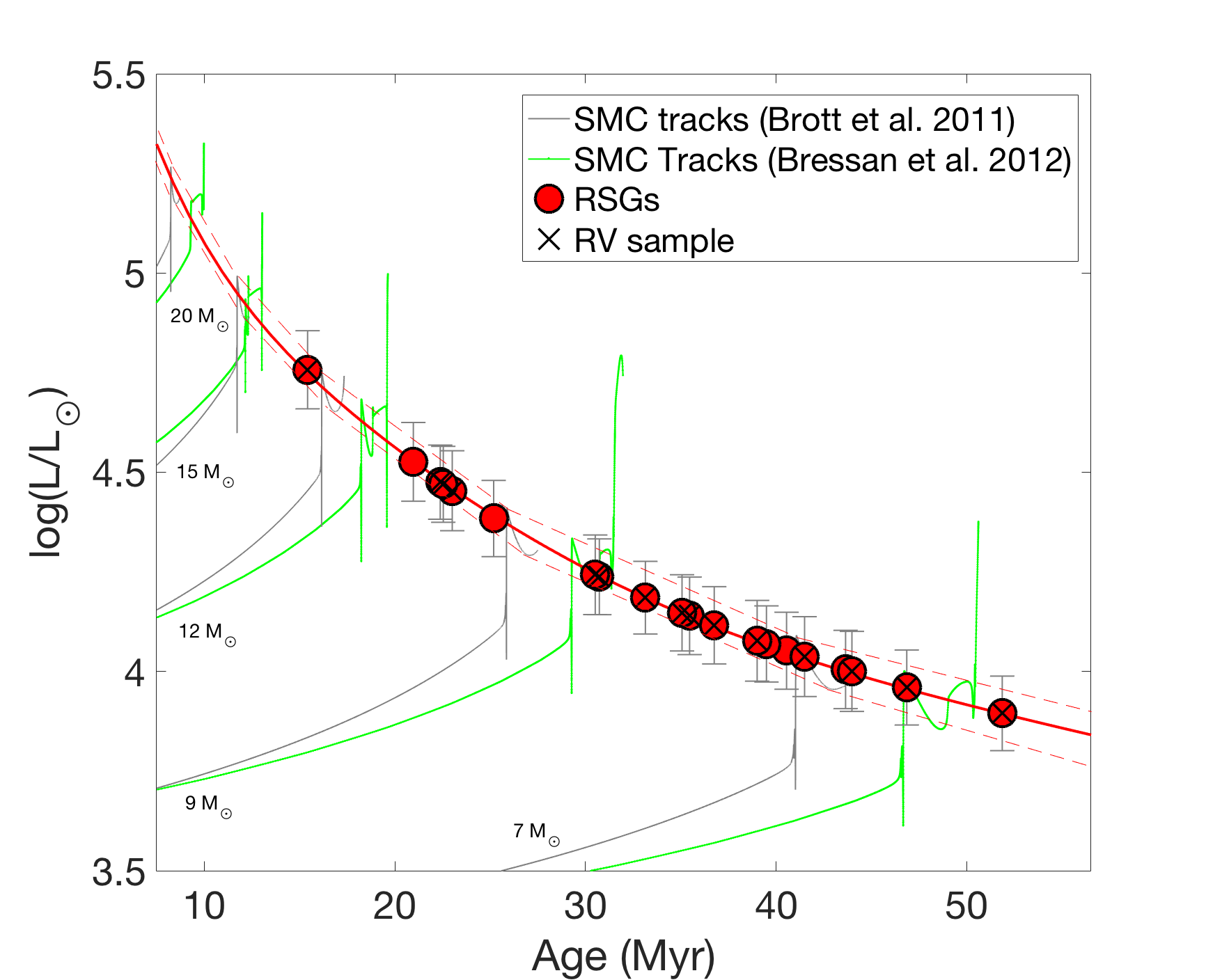}  
  \caption[]{\label{fig:ngc330LAge} %
  Luminosity-age diagram for all RSGs in NGC\,330, stars that have RV measurements from the present study are marked with a cross. Luminosities are derived the $J$-band magnitude calibration of~\citet{2013ApJ...767....3D}. Using the least luminous RSGs, we estimate the age of the NGC\,330 to be 45\,$\pm$5\,Myr. SMC-like evolutionary tracks from ~\citet{2011A&A...530A.115B} are used to estimate ages, and~\citet{2012MNRAS.427..127B} tracks are shown to highlight differences between models.}
\end{figure}

NGC\,330 is a well-observed cluster with several studies of its CMD , implying a range of potential ages from approximately 20 to 40 Myr (see Table~\ref{tb:ngc330prop} and accompanying references).  To some extent, this spread in age stems from the difficulty that theoretical isochrones have in simultaneously reproducing salient features in the CMD of this cluster, and in analogous clusters in the Milky Way and the LMC (essentially clusters hosting both BSG and RSG stars). 
These features include the main-sequence turn-off (MSTO), the main-sequence turn-on (MSTON), or pre-main-sequence stars, the mean magnitude of the BSG stars, and the mean magnitude of the RSG stars \citep{2016ApJ...833..154C}.  
In addition to the above approaches, \citet{b19} and \citet{bdsb19} have proposed a new technique for determining the ages of such clusters that uses the luminosity distribution of their RSG stars, with the former finding good agreement with the MSTON age of the LMC cluster Hodge\,301, but disagreeing with the more traditional MSTO approach.  
This new technique argues that the luminosity spread of RSGs in a cluster is the result of stellar mergers, recognising that the RSG population of any given cluster is likely affected by multiplicity on the main sequence. 
In this scenario the most luminous red straggler RSGs are the evolutionary descendants of rejuvenated massive stars on the main sequence that are the result of the merging of a binary system. 
It then follows that the distribution in luminosity of the faint portion of the RSG branch is an age indicator for a cluster. We used the age-luminosity
diagram shown in
Figure~\ref{fig:ngc330LAge} to estimate the age of NGC\,330 with this new method.

For this analysis we selected targets
based on their {\em Gaia} colours, with the criteria
$G$\,$<$\,15\,mag and $BP-RP$\,$>$\,1.2\,mag \citep[Gaia
collaboration,][]{2018A&A...616A...2L}, where the sample was first
cleaned based on their kinematic properties, as described in
Section~\ref{sec:obs}.  This yielded 20 RSGs, 15 of which are from the
HARPS + MUSE sample and a further 5 targets outside the MUSE
field of view and with no HARPS data. The luminosities of our 
RSG sample were derived using the $J$-band magnitude calibration
from \citet{2013ApJ...767....3D} with a distance modulus of 18.95~\citep{2014ApJ...780...59G}. 

We estimated the age of NGC\,330 using the 10th percentile of the luminosity distribution, resulting in an age of 45\,$\pm$5\,Myr (see Table~\ref{tb:ngc330prop} for a
summary of NGC\,330 cluster properties). We note that
there is a clear distinction (one magnitude in the {\em Gaia}
$G$ band, see Figure~\ref{fig:cmd}) between the least luminous star shown in
Figure~\ref{fig:ngc330LAge} and the population of less massive
red stars in the region. 
\citet{bdsb19} demonstrated that the
lowest luminosity RSG expected for a cluster of 40\,Myr containing 50 RSGs with
non-rotating MESA isochrones~\citep{2016ApJS..222....8D} is
$\log$(L/L$_\odot$)\,$=$\,3.9, in excellent agreement with what is
observed in NGC\,330.
Table~\ref{tb:ngc330prop} compiles the physical
properties of NGC\,330 and summarises the main results of this study.

\citet{b19} suggested to use an arbitrary cut-off of $\log$(L/L$_\odot$)\,$=$\,4.3 to distinguish between RSGs and lower-mass stars, which they argued should not be considered in the age analysis. Contamination from SMC-field stars is not likely to be significant for our sample in this respect, and we consider the exact distinction between RSG and massive-AGB stars, which are known to have an increasingly similar evolutionary path at lower metallicity~\citep{2010MNRAS.401.1453D,2015MNRAS.446.2599D}, unimportant for an analysis of the age of the cluster.
We did consider the entire sample as genuine RSGs, however.

Comparing the age derived here with previous values in the literature, see Table~\ref{tb:ngc330prop}, we find that the RSG age based on the current technique is older than that derived from more traditional approaches such as fitting the MSTO, or the mean positions of the RSG and BSG stars in the CMD, as discussed above.  

We note that recent work by \citet{2018MNRAS.477.2640M} argued for two stellar populations in this cluster, one bi-modal population of 40\,Myr, composed of a sub-population of stars rotating at close to critical rotational velocity, and a sub-population with low rotational velocity, and a second younger population of 32\,Myr with low rotational velocities. 
Figure~\ref{fig:ngc330LAge} shows that this age spread cannot explain the observed spread in apparent ages of the RSGs in the cluster, which range from 15 to 50\,Myr. 
However, spectroscopic studies of the brightest turn-off stars in NGC\,330 consistently find them to be blue stragglers, which are B-type giants beyond the end of the classical main sequence \citep{1996A&A...311..470G, 2003A&A...398..455L}. As \citet{b19} discussed, these may be the evolutionary ancestors of the bright RSGs.

\begin{table}
\caption{Physical properties of NGC\,330.}              
\label{tb:ngc330prop}      
\centering                                      
\begin{tabular}{l lc}          
\hline\hline                        
Quantity & Value & Ref.\\    

\hline                                   
  $\tau_{\rm TOFF}$       & 20\,$\pm$\,1.5\,Myr & (1)\\ 
                          & 25\,$\pm$\,15\,Myr & (2)\\ 
                          & 32\,Myr & (3)\\            
                          & 32 \& 40\,Myr & (4)\\      
                          & 35-40\,Myr & (5)\\
  $\tau_{\rm RSG}$       & 45\,$\pm$\,5\,Myr & (6)\\ 
  $\log$(M$_{\rm dyn}$/M$_\odot$)  & 5.20\,$\pm$\,0.17 & (6)\\ 
  $\log$(M$_{\rm phot}$/M$_\odot$) & 4.6\,$\pm$\,0.2 & (7)\\  
  $v_0$                   & 153.7\,$\pm$\,0.8\,\kms & (6)\\   
  $\sigma_{\rm 1D}$       & 3.20\,$^{+0.69}_{-0.52}$\,\kms & (6)\\ 
  $f_{\rm MS}$ & 0.54\,$\pm$\,0.05 & (8)\\  
  $f_{\rm RSG}$ & 0.3\,$\pm$\,0.1 & (6)\\   
\hline                                             
\end{tabular}
\tablefoot{(1) \cite{1994AAS...185.5112G}, (2) \cite{1995A&A...293..710C}, (3) \cite{2000AJ....119.1748K}, (4) \cite{2018MNRAS.477.2640M}, (5) ~\cite{2019arXiv191103477B}, (6) this study , (7) \cite{2003MNRAS.338...85M}, (8) \cite{2017ApJ...844..119L}.}
\end{table}

\subsection{Evolutionary considerations} 
\label{sub:evo}

\citet{2017ApJ...844..119L} studied the extended main-sequence turn-off
in four young massive clusters in the Magellanic Clouds (including
NGC\,330). Using {\em HST} photometry and synthetic CMDs, they estimated a main-sequence binary fraction of $f_{\rm
  MS}$\,$=$\,0.54\,$\pm$\,0.05 using the method
of~\citet{2012A&A...540A..16M} and taking into account all mass ratios
assuming a flat mass-ratio distribution. This is significantly larger
than our estimated $f_{\rm RSG}$, as expected. However, we caution that these
estimates used very different approaches and therefore have very different biases. 
A more appropriate comparison would be with a binary fraction estimated using RV variations.

The spread of luminosities among the RSG population is potentially an
indicator of post-interaction binary products, with a larger spread
indicating greater \textquoteleft contamination\textquoteright~from
binarity.  The observed luminosity spread in NGC\,330 is
characteristic of those observed in other young massive clusters in
the Galaxy and LMC and SMC~\citep[e.g.][]{2016MNRAS.458.3968P}.
Following the arguments in \citet{b19}, we assumed that the relative luminosity range of a single RSG in a coeval cluster is $\sim$5\% and therefore considered all RSGs with luminosities above $\log$L/L$_\odot$~=~4.2 to be red stragglers.
This resulte  in a red-straggler fraction in NGC\,330 of 50\%, in excellent agreement with the results of \citet{b19}. 
A more quantitative comparison is beyond the scope of this study, but if
the spread of luminosities of RSGs can be attributed to binary
post-interaction products, this would suggest similar binary fractions
within young massive clusters.

\section{Conclusion} \label{sec:conclusion}

We have investigated the multiplicity properties of the RSG population
of NGC\,330 using archival multi-epoch HARPS data supplemented with
multi-epoch MUSE observations. Precise RVs were estimated
for our targets with a slicing approach that is specifically designed
to provide an unbiased estimate of the atmospheric RV, as presented by
P19.  Based on the stability and high resolution of the HARPS
spectrograph, the estimated precision on the RVs for our RSGs is as
small as 60\,$\mbox{m s}^{-1}$.

The line-of-sight velocity and its associated dispersion were estimated
for the cluster, superseding past estimates that were limited by
observational precision \citep{1980MNRAS.191..285F}. Given their
apparent lack of RV variability, we argue that RSGs are 
effective kinematic tracers of young massive clusters. For a larger
observational sample, we combined our estimates for the RSGs with
RVs for nearby OBA-type massive stars, and the resulting velocity
dispersion ($\sigma_{\rm 1D}$\,$=$\,3.20\,$^{+0.69}_{-0.52}$\,\kms) 
is in good agreement with the value estimated from the RSGs alone.

When we assumed virial equilibrium, the dynamical mass of the cluster was
estimated as log\,(M$_{\rm dyn}$/M$_{\odot}$)\,$=$\,5.20\,$\pm$\,0.17, in
good agreement with previous dynamical mass estimates and photometric mass estimates from
\citet{2003MNRAS.338...85M} and~\citet{2019arXiv191103477B},
assuming a binary fraction of 25\% using the relations of~\citet{2010MNRAS.402.1750G}.

The multiplicity properties of the RSGs were investigated by searching
for significant variability within the multi-epoch RVs for each RSG.
By varying the criteria to detect significant variability, we attempted
to distinguish between variations from atmospheric motions and genuine
binarity.  To estimate the intrinsic binary fraction of the RSG
population, we simulated binary populations using an empirically defined
distribution of systems from~\citet{2017ApJS..230...15M}.  For orbital
periods in the range 2.3\,$<$\,log\,P\,[days]\,$<$\,4.3 and mass ratios above 0.1,
we estimate an intrinsic binary fraction of $f_{\rm RSG}$~=~0.3\,$\pm$\,0.1.
We compile the physical properties of NGC\,330 in Table~\ref{tb:ngc330prop} and show that the binary fraction of RSGs appears significantly smaller than that of main-sequence stars, which is expected assuming that binary interactions remove RSG binaries.

Based on a sample of 20 RSGs, selected through the Gaia DR2 astrometry (15 of which have RVs presented in the current study), the age of NGC\,330 is estimated using the distribution of luminosities of the RSG population of NGC\,330 following~\citet{b19}.
This technique takes into account the affect of binarity on the RSG population, and we estimate the cluster age to be $\tau_{\rm RSG}$~=~45\,$\pm$\,5\,Myr. We note that the suggestion of \citet{2018MNRAS.477.2640M}, that the main-sequence population of NGC\,330 consists of 32\,Myr and 40\,Myr populations, cannot explain the spread in luminosities that is observed in the RSG population.
Using the luminosity distribution, we estimate a red straggler fraction of 50\%, in excellent agreement with \citet{b19}.

\begin{acknowledgements}

The authors would like to thank the anonymous referee
for a careful review that improved the article.
L. R. P and A. H acknowledge support from grant AYA2015- 68012-C2- 1-P from the Spanish Ministry of Economy and Competitiveness (MINECO).
This research is partially supported by the Spanish Government under grants AYA2015-68012-C2-1/2-P and PGC2018-093741-B-C21/2 (MICIU/AEI/FEDER, UE).
This publication makes use of data products from the Two Micron All Sky Survey, which is a joint project of the University of Massachusetts and the Infrared Processing and Analysis Center/California Institute of Technology, funded by the National Aeronautics and Space Administration and the National Science Foundation.
This work has made use of data from the European Space Agency (ESA) mission
{\it Gaia} (\url{https://www.cosmos.esa.int/gaia}), processed by the {\it Gaia}
Data Processing and Analysis Consortium (DPAC,
\url{https://www.cosmos.esa.int/web/gaia/dpac/consortium}). Funding for the DPAC
has been provided by national institutions, in particular the institutions
participating in the {\it Gaia} Multilateral Agreement.
We acknowledge support from the FWO-Odysseus program under project G0F8H6N.
This project has further received funding from the European Research Council under European Union's Horizon 2020 research programme (grant agreement No 772225 - MULTIPLES). TOPCAT~\citep{2005ASPC..347...29T} was used to handle the {\em Gaia} data used in this publication.
This research made use of Astropy,\footnote{http://www.astropy.org} a community-developed core Python package for Astronomy \citep{2013A&A...558A..33A,2018AJ....156..123A}

\end{acknowledgements}

\bibliographystyle{aa} 
\bibliography{journals}      

\begin{appendix}
\section{Multi-epoch radial velocities}
Table~\ref{tb:rvall} provides the complete list of estimated 
RVs for the RSGs considered in this paper as well as literature measurements.

\begin{figure*}[hbtp]
  \centering
  \includegraphics[width=\linewidth]{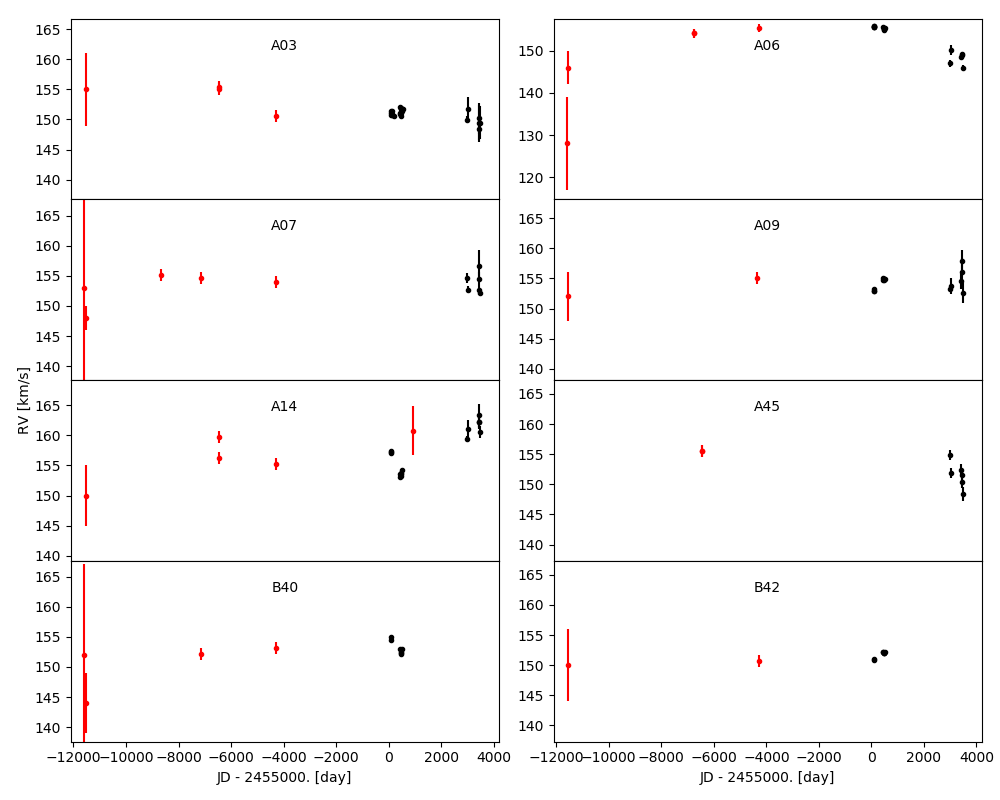}  
  \caption[]{\label{fig:rv_lit} %
  RV estimates for all targets with literature measurements, shown on the same relative scale to demonstrate the observed differences in RV variations (except for Rob74~B31, which will be discussed in detail in an upcoming publication). Red data points indicate RV estimates from literature studies. 
  In general, the agreement with the literature measurements is excellent. 
}
\end{figure*}

Figure~\ref{fig:rv_lit} displays RV measurements for our targets alongside literature measurements spanning a baseline of more than 40\,years. 
This figure highlights the excellent agreement between the RV measurements presented in this study with those published previously, and simultaneously highlights the lack of significant RV variability over this timescale.
However, because it is difficult to determine the systematic errors that might be present in these previous results, we did not include these data in our quantitative determination of the binary fraction. Instead, we provide a brief qualitative discussion of these results here.

 ~\citet{1980MNRAS.191..285F} published some of the earliest multi-epoch RVs for the brightest stars in NGC\,330, finding an unusually small fraction of  variables that they interpreted as evidence for a small binary fraction in the cluster. Their measured values for several RSGs in the cluster show little dispersion per star, with an observational uncertainty per star of $\sim \pm 1-3$ km\ s$^{-1}$. There is an indication of bimodality in their results, however, that suggests some unresolved systematic uncertainty for some stars of $\sim$5 km\, s$^{-1}$.
 \citet{1985AJ.....90.1196C} measured RVs for many stars in the cluster with similar mean results as \citet{1980MNRAS.191..285F}, but with uncertainties of $\sim$10 km\, s$^{-1}$ for each star. 
  The high-resolution echelle observations of \citet{1991A&A...252..557S}, \citet{1999A&A...345..430H} and \citet{1999AJ....117.2286G} have quoted uncertainties of 1 km/s, derived from cross correlation with standards, and compare extremely well with the HARPS results presented here (but for some observations, only approximate dates are given).
These echelle observations are remarkably constant, within the small uncertainties, and are all consistent with the systemic RV of the cluster. The overall picture therefore is that the RSG RVs discussed in this paper are remarkably constant over a time span of $\sim 40$ years.
\begin{table*}
\caption{RV estimates at each epoch for RSGs in the HARPS and MUSE samples.}\label{tb:rvall}
\centering                                      

\begin{tabular}{lcccl}          
\hline\hline      
ID & JD & RV (\kms) & $\sigma_{\rm RV}$ (\kms) & Spectrograph/Reference \\

\hline
A3  &  2443462.5 &  155.0 &  6   &  1980MNRAS.191..285F \\
A3  &  2450728.5 &  150.5 &  1   &  1999AJ....117.2286G \\
A3  &  2448544.5 &  155.4 &  1   &  1999A\&A...345..430H \\
A3  &  2448544.5 &  155.1 &  1   &  1999A\&A...345..430H \\
A3  & 2455087.7 & 150.77 & 0.19 & HARPS  \\
A3  & 2455088.6 & 151.39 & 0.17 & HARPS  \\
A3  & 2455100.7 & 151.10 & 0.24 & HARPS  \\
A3  & 2455120.7 & 151.37 & 0.15 & HARPS  \\
A3  & 2455144.5 & 150.99 & 0.15 & HARPS  \\
A3  & 2455145.6 & 150.70 & 0.30 & HARPS  \\
A3  & 2455185.6 & 150.52 & 0.20 & HARPS  \\
A3  & 2455430.7 & 150.89 & 0.17 & HARPS  \\
A3  & 2455432.6 & 151.00 & 0.34 & HARPS  \\
A3  & 2455447.6 & 152.05 & 0.09 & HARPS  \\
A3  & 2455467.6 & 151.15 & 0.17 & HARPS  \\
A3  & 2455470.5 & 150.54 & 0.15 & HARPS  \\
A3  & 2455477.5 & 151.82 & 0.08 & HARPS  \\
A3  & 2455479.6 & 151.89 & 0.20 & HARPS  \\
A3  & 2455502.5 & 151.43 & 0.16 & HARPS  \\
A3  & 2455535.6 & 151.80 & 0.16 & HARPS  \\

\ldots & \ldots & \ldots & \ldots & \ldots \\
\hline                                             
\end{tabular}
\tablefoot{A full version of this table is available electronically. The first 20 lines are shown as a sample.
}
\end{table*}

\section{Literature IDs}
Compiling the archival data used in this study, we realised that owing to the dense cluster environment, identifying the targets in different literature studies is often difficult. To aid future studies, here we cross-match the identifications of several well-known studies of NGC\,330 with {\em Gaia} DR2 identifications. Table~\ref{tb:IDs} lists the results including {\em Gaia} astrometry for the sources.

\begin{table*}
\caption{Identifications for RSGs in NGC\,330 within a 2\farcm5 radius of the cluster centre.}
\label{tb:IDs}      
\centering                                      
\begin{tabular}{lccccc}          
\hline\hline                        
ID$_{R74}$ & ID$_{Arp59}$ & ID$_{Bal92}$ & ID$_{Gaia DR2}$& RA & DEC\\    
\hline
A3             & ARP26 &  BAL690   & 4688993681762966400 &  00:56:21.6996 & $-$72:27:35.8594 \\
A6             & ARP43 &  BAL450   & 4688993647403231232 &  00:56:13.8954 & $-$72:27:32.4019 \\
A7             & ARP41 &  BAL475   & 4688993647403173504 &  00:56:14.5617 & $-$72:27:42.4507 \\
A9             & ARP42 &  BAL423   & 4688993647403241088 &  00:56:12.8613 & $-$72:27:46.6309 \\
A14            & ARP17 &  BAL623   & 4688993471264541056 &  00:56:19.0367 & $-$72:28:08.2365 \\
A27            & ARP32 &  BAL561   & 4688993681762883584 &  00:56:17.0829 & $-$72:27:35.0108 \\
A42           & ARP21 &  BAL645   & 4688993681762909440 &  00:56:20.1423 & $-$72:27:47.0941 \\
A45            & ARP38 &  BAL618   & 4688993681762919040 &  00:56:18.8279 & $-$72:27:47.4984 \\
A46            & ARP37 &  BAL595   & 4688993677426863872 &  00:56:18.0162 & $-$72:27:49.8524 \\
A52            & ARP36 &    --     & 4688993681703883648 &  00:56:17.3709 & $-$72:27:52.5296 \\
A57            & ARP19 &  BAL619   & 4688993681762914304 &  00:56:19.4722 & $-$72:27:53.2203 \\
B15            & ARP29 &  BAL523   & 4688993750482336640 &  00:56:16.5574 & $-$72:27:04.6490 \\
B31            & ARP02 &  BAL317   & 4688993643141502336 &  00:56:09.4247 & $-$72:28:09.4895 \\
B40            & ARP10 &  BAL801   & 4688993475604563584 &  00:56:25.3298 & $-$72:28:18.2008 \\
B42            & ARP12 &  BAL878   & 4688993471342994048 &  00:56:27.3354 & $-$72:28:06.4611 \\
B10      & ARP III-224 &  --       & 4688999557280890496 &  00:56:36.3522 & $-$72:26:46.4698 \\
B19      & ARP III-205 &  --       & 4688994368957731584 &  00:55:55.7677 & $-$72:27:09.8270  \\
B20      & ARP III-203 &  BAL90    & 4688993716122719872 &  00:56:00.9736 & $-$72:27:39.6590  \\
--       & ARP III-210 &  --       & 4688994472036827392 &  00:56:10.9903 & $-$72:26:26.0654 \\
--       & ARP III-214 &  --       & 4688994575116023808 &  00:56:14.7713 & $-$72:25:20.7020  \\
\hline                                             
\end{tabular}
\tablefoot{Literature identification from \cite{1974A&AS...15..261R} and \cite{1992MNRAS.256..425B}.}
\end{table*}

\end{appendix}

\end{document}